\documentclass[aps,pre,showpacs,amsmath,amssymb]{revtex4}
\usepackage{graphicx}

\newcommand{\bs}[1]{{\boldsymbol #1 }}
\newcommand{\ave}[1]{\langle #1 \rangle}

\usepackage{graphicx}
\usepackage{bm}

\begin{document}

\preprint{}

\title{ Current in coherent quantum systems connected to mesoscopic Fermi reservoirs}

\author{Shigeru Ajisaka}
\affiliation{Departamento de F\'isica, Facultad de Ciencias F\'isicas y Matem\'aticas, Universidad de Chile, Casilla 487-3, Santiago Chile}
\author{Felipe Barra}
\affiliation{Departamento de F\'isica, Facultad de Ciencias F\'isicas y Matem\'aticas, Universidad de Chile, Casilla 487-3, Santiago Chile}
\author{Carlos Mej\'ia-Monasterio}
\affiliation{Laboratory of Physical Properties, Technical University of Madrid, Av. Complutense s/n 28040, Madrid, Spain}
\author{Toma\v{z} Prosen}%
\affiliation{Faculty of Mathematics and Physics, University of Ljubljana, Jadranska 19, SI-1000 Ljubljana, Slovenia}

\date{\today}
\begin{abstract}
  We study particle current in a recently proposed model for coherent quantum transport.
In this model a system connected to mesoscopic Fermi reservoirs (meso-reservoir) is driven out of
equilibrium by the action of super reservoirs thermalized to prescribed temperatures and
 chemical potentials by a simple dissipative mechanism described by the
Lindblad equation.  We compare exact (numerical) results with theoretical expectations based on the Landauer formula. 
\end{abstract}


\maketitle


\section{ Introduction}

Particle current trough a coherent mesoscopic conductor connected at its left and right hand side to reservoirs is usually
described in the non-interacting case by a formula, due to Landauer, based on the following physical picture:   
Electrons in the left (right) reservoir which are Fermi distributed with chemical potential $\mu_{L}$ ($\mu_{R}$) and inverse temperature $\beta_{L}$ ($\beta_{R}$) can come close to the conductor and feed a scattering state that can transmit it to the right (left) reservoir. All possible dissipative processes such as thermalization occur in the reservoirs while the system formed by the conductor and the leads is assumed to be coherent. 
The probability of being transmitted is  a property of the conductor connected to the leads which is treated as a scattering system. 
In this picture, the probability that an outgoing electron comes back to the conductor before being thermalized is neglected, 
the contact is said to be reflectionless. 
This description of the non-equilibrium steady state (NESS) current through a finite system 
has been rigorously proved in some particular limiting situations~\cite{needref}, such as infinite reservoirs, 
but several difficulties prevail the understanding of non-equilibrium states in general and
the description of current in more general situations, for instance in the case of interacting particles.  
Two frameworks are usually considered to study these open quantum systems: 
one deals with the properties of the state of the total (infinite) system
\cite{Rue00,tasaki} where reservoirs are explicitly considered as part of the system.
The other is based on the master equation of the reduced density operator,
obtained by tracing out the reservoirs'
degrees of freedom and is better suited to be applied to different systems and to compute explicitly 
some averaged NESS properties, at the price of several approximations such
as {\it e.g.}, Born-Markov (see e.g.  \cite{Wichterich07}).

In this paper we explore particle current in a model where we mimic the leads that connect the reservoirs with the system,
as a finite non-interacting system with a finite number of levels (which we call meso-reservoir). The reservoirs (called here super-reservoirs) are modeled by local Lindblad operators representing the effect that a Markovian macroscopic reservoirs have over the meso-reservoirs. In Sec.~\ref{sec.2} we introduce the model and briefly review the method we use to solve it. In Sec.~\ref{sec.3} we analyze the particle current operator and indicate the quantities that should be computed
for a full description of the current. In Sec.~\ref{Landauer} we briefly present the Landauer formula that is expected to apply in some appropriate limits to our model and in Sec.~\ref{numeric} we analyze the numerical results we obtained with our model and compare them with the current predicted by the Landauer formula, validating the applicability of our model but also going beyond by computing the full probability distribution function (PDF) of the current. In Sec.~\ref{conclu} we present some conclusions and discuss interesting perspectives of our study.

\section{Description of the model} 
\label{sec.2}
We consider a one-dimensional quantum chain of spinless fermions coupled at its boundaries to meso-reservoirs comprising a finite number of spinless fermions with wave number $k$ ($k\in\{1,\ldots,K\}$). 
The Hamiltonian of the total system can be written
as $H = H_S+H_L+H_R+V$, where
\begin{equation}
\label{hs}
H_S = -\sum_{j=1}^{N-1} \Big(t_j c_j^\dag c_{j+1} + ({\rm h.c.}) \Big)
+\sum_{j=1}^{N} U c_j^\dag c_j
\end{equation}
is the Hamiltonian of the chain 
with $\{t_j\}$ the nearest neighbor hopping, $U$ the
onsite potential and  $c_j,c_j^\dagger$ the  annihilation/creation operator for the spinless fermions on the site $j$ of the chain (conductor). The chain interacts through the term
\begin{equation}
V =\sum_{k=1}^{K} \left( v^L_{k}a_{kL}^\dag c_1 + v^R_{k}a_{kR}^\dag c_n\right) + ({\rm h.c.}),
\label{V1}
\end{equation}
with the meso-reservoirs $H_{\alpha}= \sum_{k=1}^{K} 
\varepsilon_k a_{k\alpha}^\dag a_{k\alpha}$.  Here $\alpha\!=\!\{L, R\}$ denotes the left and right meso-reservoir.
They share the same spectrum with a constant density of states $\theta_0$ in the band $[E_{\rm min},E_{\rm max}]$ described by $\varepsilon_k \equiv \theta_0 (k-k_0)$ and  
$a_{k,\alpha},a^\dagger_{k,\alpha}$  are annihilation/creation operator  of the  left  and right meso-reservoirs.
The system is coupled to the leads only at the extreme sites 
of the chain with coupling strength $v_k^\alpha$ that we choose $k$-independent~\footnote{In general we can include couplings to deeper sites of the chain and also $k$ dependent super-reservoir to meso-reservoir couplings $\gamma_k$} $v_k^\alpha = v_\alpha$.

We assume that  the density matrix of the chain - meso-reservoirs system evolves according to the 
many-body Lindblad equation
\begin{eqnarray}
\frac{{\rm d}}{{\rm d}t}\rho &=&
-i[H,\rho] 
+\sum_{k,\alpha,m}
\left(
2L_{k, \alpha, m} \rho L_{k, \alpha, m}^\dag
-\{L_{k, \alpha, m}^\dag L_{k, \alpha, m},\rho\}
\right),
\label{eq:lindbladmodel}
\end{eqnarray}
where $m\in\{1,2\}$ and $L_{k,\alpha,1} = \sqrt{\gamma ( 1 - F_{\alpha}(\varepsilon_k) )} a_{k\alpha},\quad L_{k,\alpha,2}=\sqrt{\gamma F_{\alpha} (\varepsilon_k)} a_{k\alpha}^\dag$ are operators  representing the coupling of the meso-reservoirs to the
super-reservoirs, $F_\alpha(\varepsilon)=(e^{\beta_\alpha  (\varepsilon-\mu_\alpha)}+1)^{-1}$   
are Fermi distributions, with inverse   temperatures $\beta_\alpha$ and chemical potentials
$\mu_\alpha$, and  $[\cdot , \cdot]$  and $\{ \cdot ,  \cdot\}$ denote
the commutator and anti-commutator, respectively. 
The parameter $\gamma$ determines the strength of the coupling to the
super-reservoirs and to keep the model as simple as possible we take it
constant.  
The form of the Lindblad dissipators is such that in the absence of coupling to the chain (i.e. $v_\alpha=0$), when the meso-reservoir is only coupled to the super-reservoir, the former is in an equilibrium state described by Fermi distribution.~\cite{prosen08,Kosov11}.

To analyze our model we use the formalism developed in
\cite{Prosen10}. There it is shown that the spectrum of the evolution superoperator is given
in terms of the eigenvalues $s_j$ (so-called rapidities) of a matrix $X$ which in our case is given by
\begin{eqnarray*}
\bs{X} &=& -\frac{\rm i}{2}\bs{H}\otimes \sigma_y + 
	\frac{\gamma}{2}
	\begin{pmatrix}
	\bs{E}_{K} & \bs{0}_{K\times N} &\bs{0}_{K\times K}\\
	\bs{0}_{N\times K} & \bs{0}_{n\times N} &\bs{0}_{N\times K}\\
	\bs{0}_{K\times K} & \bs{0}_{K\times N} &\bs{E}_{K}
	\end{pmatrix}
	\otimes\bs{E}_2,
\end{eqnarray*}
where $\bs{0}_{i\times j}$ and $E_j$ denote $i\times j$ zero matrix
and $j\times j$ unit matrix, $\sigma_y$ is the Pauli matrix, and
$\bs{H}$ is a matrix which defines the quadratic form of the
Hamiltonian, as $H= \bs{d}^\dagger \bs{H} \bs{d}$ in terms of fermionic
operators $\bs{d}^T \equiv \{a_{1L},\cdots
a_{KL},c_1,\cdots,c_N,a_{1R},\cdots a_{KR} \}$.

The NESS average of a quadratic observable like $d^\dagger_jd_i$ is given~\cite{Prosen10} in terms of the
solution $\bs{Z}$ of the Lyapunov equation $\bs{X}^T \bs{Z}+\bs{Z}\bs{X} \equiv  \bs{M}_i$ with $\bs{M}_i \equiv -\frac{i}{2} {\rm diag}\{m_{1L},\cdots,m_{KL},
\bs{0}_{1\times N},\, m_{1R},\cdots,m_{KR}\}
\otimes \bs{\sigma_y}$ and $m_{k\alpha}=\gamma\{ 2F_\alpha (\varepsilon_k)-1 \}$ as follows: Consider the change of variables $w_{2j-1}\equiv d_j+d_j^\dag,\ \ w_{2j}\equiv i(d_j-d_j^\dag)$, the NESS average of the quadratic observable $w_j w_k$ is determined by the matrix $\bs{Z}$ through the relation $\ave{ w_j w_k}=\delta_{j,k} -4i\bs{Z}_{j,k}$.
Wick's theorem can be used to obtain expectations of
higher-order observables and in fact, the full probability distribution for these
expectation values in some cases.

\section{Particle current}
\label{sec.3}
The operator representing the current flowing from the $k$-th level of the meso-reservoir to the chain is given by 
 \begin{equation}
 j^L_{k}=iv^L_{k}(a_k^\dagger c_1-c_1^\dagger a_k),
 \end{equation}
while the current trough the site $l$ of the chain is
\begin{equation}
J_l=it_l(\hat{c}_l^\dag \hat{c}_{l+1}- \hat{c}_{l+1}^\dag \hat{c}_l).
\end{equation}
In the steady state the average current is conserved in this model~\cite{mariborg,Shigeru} and thus $\langle J_l\rangle $ is independent of $l$. Moreover, if we define the current from the left meso-reservoir as $ J=\sum_{k=1}^K j^L_k$, we have that $\langle J_l\rangle =\langle J\rangle $. 

It is not difficult to note that current 
satisfies
\begin{equation}
J_l^n=\left\{
\begin{array}{cc}
t_l^{n-1}J_l & \,{\rm if} \,n \,{\rm odd,}\\
t_l^{n-2}J_l^2 &  \,{\rm if} \,n \,{\rm even}\\  
\end{array}
\right.
\label{moments}
\end{equation}
with $J_l^0=1.$
Now we are in a position to compute the full non equilibrium current distribution in terms of $\langle J_l\rangle$ and $\langle J_l^2\rangle$. For this we consider the generating function
\begin{equation}
\langle e^{ikJ_l}\rangle=\sum_{n=0}^\infty \frac{(ik)^n}{n!}\langle J_l^n\rangle,
\end{equation}
which, using Eq.(\ref{moments}), gives
\begin{equation}
\langle e^{ikJ_l}\rangle=\left(1-\frac{\langle J_l^2\rangle}{t_l^2}\right)+\frac{\langle J_l^2\rangle}{t_l^2}\cos kt+
\frac{\langle J_l\rangle}{t_l}i\sin kt.
\end{equation}
The probability distribution $p(J_l)$ is the inverse Fourier transform of $\langle e^{ikJ_l}\rangle$, thus we get
\begin{equation}
p(J_l)=\left(1-\frac{\langle J_l^2\rangle}{t_l^2}\right)\delta(J_l)+
\left(\frac{\langle J_l^2\rangle}{2t_l^2}+\frac{\langle J_l\rangle}{2t_l}\right)\delta(J_l-t_l)+
\left(\frac{\langle J_l^2\rangle}{2t_l^2}-\frac{\langle J_l\rangle}{2t_l}\right)\delta(J_l+t_l),
\end{equation}
which is normalized.

%
We note that normality and positivity of probability lead an interesting inequality: $\frac{\langle J_l\rangle}{2t_l} <\frac{\langle J_l^2\rangle}{2t_l^2}<1-\frac{\langle J_l\rangle}{2t_l}.$ 
An equivalent result holds for the current from the $k$-level of the meso-reservoir to the system: 
\begin{equation}
p(j_{k}^L)=\left(1-\frac{\langle (j^{L}_{k})^2\rangle}{(v^{L}_{k})^2}\right)\delta(j^L_{k})+
\left(\frac{\langle (j^{L}_{k})^2\rangle}{2(v^{L}_{k})^2}+\frac{\langle j^L_{k}\rangle}{2v^L_{k}}\right)\delta(j^L_{k}-v^L_{k})+
\left(\frac{\langle (j^{L}_{k})^2\rangle}{2(v^{L}_{k})^2}-\frac{\langle j^L_{k}\rangle}{2v^L_{k}}\right)\delta(j^L_{k}+v^L_{k}).
\end{equation}

These expressions are expected because $\pm t_l$ and $0$ are the possible eigenvalues of the operator $J_l$ (similarly for $j_{k}^L$), but they show that $\langle J_l\rangle$ and $\langle J_l^2\rangle$ contains all the information about the current. We will study these quantities numerically, thus we need to solve the above mentioned Lyapunov equation and note that in the $w_j$ variables (we use the primed variables to indicate indices as they appear in variable ${\bf d}$, i.e. if $j$ is a site in the chain, then $j'=K+j$)
\begin{equation}
J_l=i\frac{t_l}{2}(
w_{2l'-1}w_{2l'+1}+w_{2l'}w_{2l'+2} ),
\end{equation} 
thus
\begin{equation}
\langle J_j\rangle=2t_j (\mathbf{Z}_{2j'-1,2j'+1}+\mathbf{Z}_{2j',2j'+2})
\end{equation}
and 
\begin{equation}
J_j^2=\frac{t_j^2}{2} (1+w_{2j'-1}w_{2j'}w_{2j'+1}w_{2j'+2} ).
\end{equation}
Wick's theorem implies
\begin{equation}
\langle J_j^2\rangle=\frac{t_j^2}{2} \big(1
-16(\mathbf{Z}_{2j'-1,2j'+2}\mathbf{Z}_{2j',2j'+1} - \mathbf{Z}_{2j'-1,2j'+1}\mathbf{Z}_{2j',2j'+2} + \mathbf{Z}_{2j'-1,2j'}\mathbf{Z}_{2j'+1,2j'+2})
\big).
\end{equation}

To simplify the discussion, in what follows we reduce the number of parameters by assuming
constant hopping and onsite energy $t_l = t$ and $U_l = U$.
Moreover we fix $\gamma$ such that $\gamma >v_{\alpha}$. In that case, we showed \cite{Shigeru} that the transport quantities become roughly independent of $\gamma$. In fact, in that case, the coupling super-reservoir -- meso-reservoir is stronger than that of the system (chain) -- meso-reservoir. Then the meso-reservoir are driven to  a near equilibrium state weakly dependent of $\gamma$. We explore now the behavior of the current as a function of $\theta_0$, $v_{\alpha}$, $t$ and contrast the observation with expectations based on the Landauer formula. Qualitative explanation of the current behavior is also provided. 

\subsection{The Landauer Formula}
\label{Landauer} 
The Landauer formula~\cite{book} provides an almost explicit expression for the NESS average current as a function of the parameters of the system. In units where $e=1$ and $\hbar=1$ it reads
\begin{equation}
\label{eq.landauer}
\ave{J}=\frac{1}{2\pi}\int d\omega(f_L(\omega)-f_R(\omega))T(\omega),
\end{equation}
where 
$
T(\omega)=
{\rm tr}[\Gamma_L(\omega)G^+ (\omega)\Gamma_R(\omega) G^-(\omega)]\,
$
is the transmission probability written here in terms of 
\begin{equation}
G^\pm(\omega)=\frac{1}{\omega-H_S-\Sigma^\pm_L(\omega)-\Sigma^\pm_R(\omega)}, 
\label{green}
\end{equation}
the retarded and advanced Green function of the system connected to the leads and of $-\Gamma_\alpha/2$ the imaginary part of the self-energy $\Sigma^+_\alpha$.

The self energies $\Sigma^\pm_\alpha$ have only terms at the boundaries of the chains~\cite{book} i.e., $(\Sigma^\pm_\alpha)_{nm}=\sigma^\pm_\alpha\delta_{nm}\delta_{n,b}$, where $b=1$ if $\alpha=L$ and  $b=N$ if $\alpha=R$ and
\begin{equation}
\sigma^\pm_\alpha=v_\alpha^2\sum_{k=1}^K\frac{1}{\omega-\varepsilon_k\pm i0}=\Lambda_\alpha(\omega)\mp\frac{i}{2}\Gamma_\alpha(\omega).
\end{equation}
Recalling that both leads have the same spectrum, we assume a constant density of lead states $1/\theta_0$ in the range $[E_{\rm min},E_{\rm max}]$, thus $\Gamma_\alpha(\omega)=2\pi v_\alpha^2/\theta_0$ is independent of $\omega$ inside the interval and zero otherwise. For the real part of the self-energy we have the principal value integral
\begin{equation}
\Lambda_\alpha(\omega)=\frac{1}{2\pi}{\rm P}\int\frac{\Gamma_\alpha(\varepsilon) d\varepsilon}{\omega-\varepsilon}=\frac{v_\alpha^2}{\theta_0} \ln\left|\frac{\omega-E_{\rm min}}{\omega-E_{\rm max}}\right|.
\end{equation}
The Landauer formula is expected to hold when the leads have a dense and wide spectrum. Therefore, 
we restrict ourselves to the case that $E_{\rm min}\ll -t$ and $t\ll E_{\rm max}$, the so called wide-band limit, where $ \Lambda_\alpha(\omega)$ can be neglected.
The transmission coefficient is then $T(\omega)=\Gamma_L \Gamma_R |G^+_{1N}(\omega)|^2$ and we need to compute the wide-band limit retarded Green function
\begin{equation}
G^+(\omega)=
\left(\begin{array}{ccccc}
\omega+i\frac{\Gamma_L}{2} & -1 & 0 & \cdots & 0\\
-1 & \omega & -1 & 0 & \vdots \\
0&-1 &\ddots & & \\
\vdots& & & & \\
0& \cdots & 0&-1&\omega+i\frac{\Gamma_R}{2}
\end{array}\right)^{-1}.
\end{equation}
Note that in the previous expression we have set $U=0$ which sets the energy axis origin and $t=1$ which sets the energy scale. Thanks to a recursion relation, this matrix can be inverted~\cite{silly} and one explicitly finds the relevant element of the Green function
\begin{equation}
\begin{array}{cc}
[G_{1N}(\omega)]^{-1}=\left(\omega +\frac{i \Gamma _L}{2}\right) \left(\omega +\frac{i \Gamma_R }{2}\right) \sum
   _{k=0}^{\left\lfloor \frac{N-2}{2}\right\rfloor } (-1)^k \omega ^{N-2-2 k}
   \binom{N-2-k}{k}\\
  -\left(2 \omega +\frac{1}{2} i (\Gamma_L +\Gamma_R )\right) \sum
   _{k=0}^{\left\lfloor \frac{N-3}{2}\right\rfloor } (-1)^k \omega ^{N-3-2 k}
   \binom{N-3-k}{k}+\sum _{k=0}^{\left\lfloor \frac{N-4}{2}\right\rfloor } (-1)^k
   \omega ^{N-4-2 k} \binom{N-4-k}{k},
\end{array}
\label{G1N}
\end{equation} 
where $\left\lfloor x\right\rfloor$ is the largest integer smaller than $x$. 
In the next section we compute numerically the integral in Eq.(\ref{eq.landauer}) and compare with the results obtained in our model.

\subsection{Numerical results}
\label{numeric}
\begin{figure}
 \includegraphics[width=7cm]{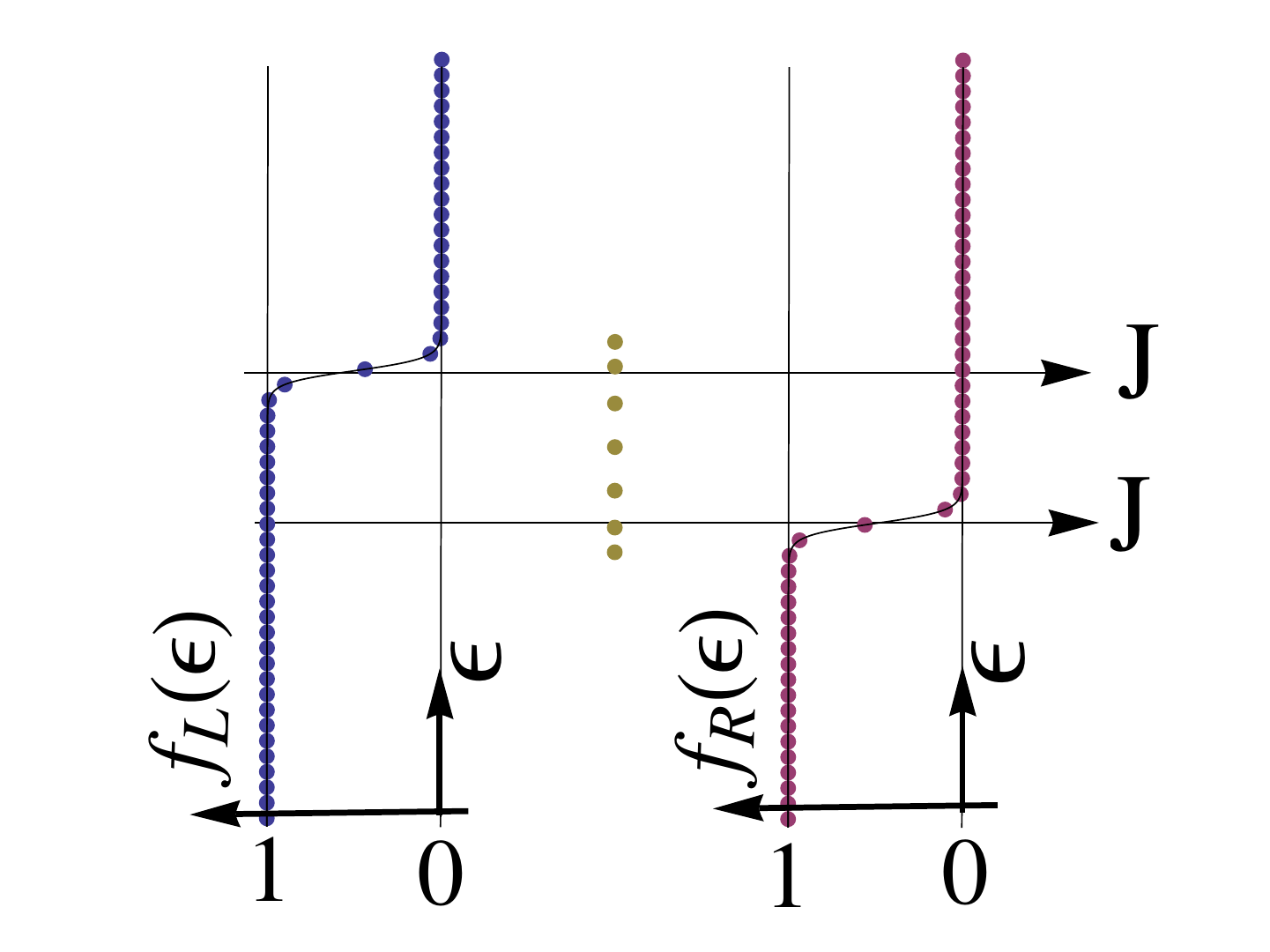}
\caption{Schematic plot of the left and right Fermi distributions and the levels of the middle chain with $N=7$ sites. }
\label{Esquema}
\end{figure}
\begin{figure}
 \includegraphics[width=7cm]{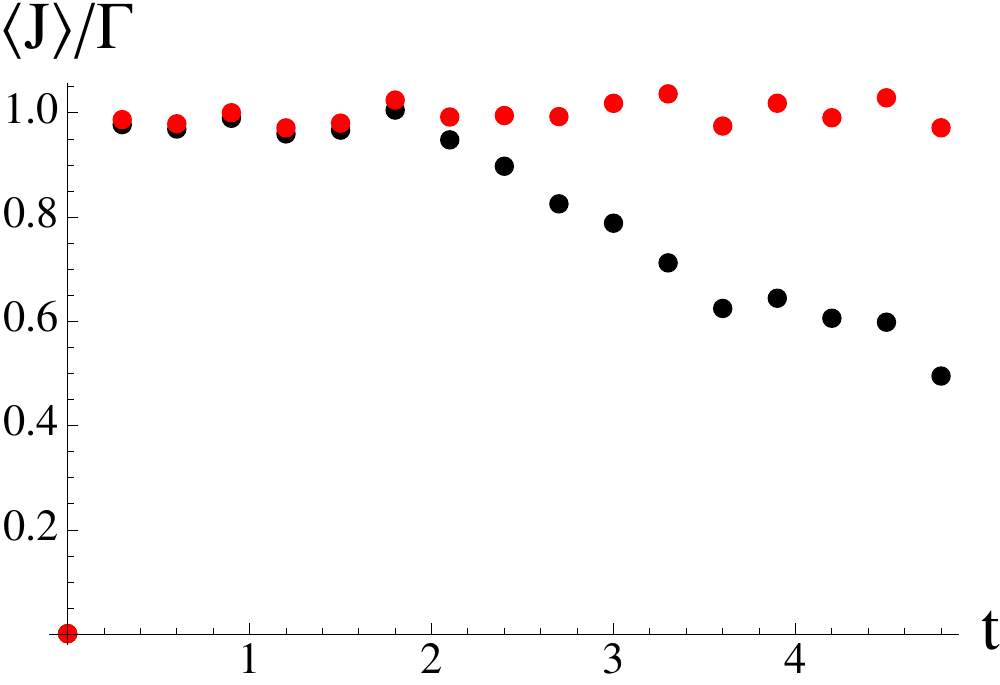}
  \includegraphics[width=7cm]{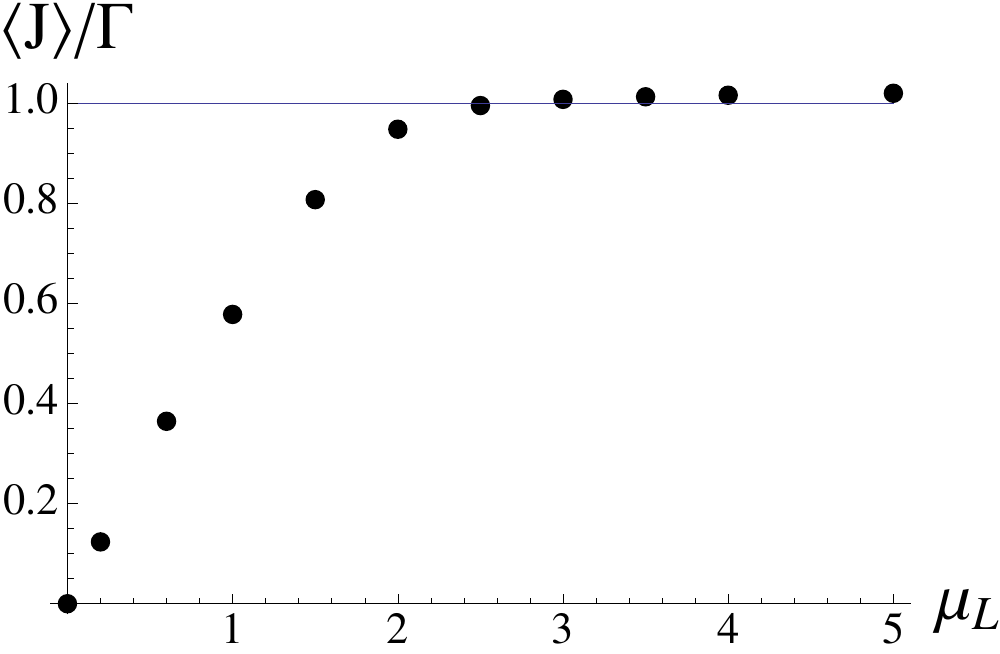}
\caption{Current versus hopping $t$ for $\gamma = 0.1,\,U=0$, $\beta_L=\beta_R=5$ , $v_L=v_R=0.05$, $K=200$ and $N=10$. In the left panel (black) $\mu_L=4,\ \mu_R=-4$ and (red) $\mu_L=10,\ \mu_R=-10$. Right panel: Same parameters except: $t = 1,\, v_L =v_R= 0.03$.}
\label{j-t}
\end{figure}

In Fig.~\ref{Esquema} we depict in blue and red two Fermi distributions with $K=50$ levels and parameters $\mu_{L}=4, \beta_L=3$ and   $\mu_{R}=-4, \beta_R=3$ respectively. In the middle (brown) the spectrum of a chain with $7$ sites and $t=3. $ The width of the chain energy band is $4t$ with $7$ levels inside and is centered around $U=0$. 
From this picture, we expect that decreasing  the width $\Delta\mu$ of the populated
energy  interval $[\mu_R,\mu_L]$  is equivalent  to  increase the
width of the  conduction band $t$. This is  confirmed in the left
panel  of  Fig.~\ref{j-t}  where  we show  that  the  current is  roughly
independent  of $t$  for $2t<\mu_L$  and decreases  with  $t$ for
$2t>\mu_L$ (black dots), when  the conduction band extends beyond
the  region populated by  electrons in  the reservoirs.   The red
dots  are  obtained  for   a  larger  $\Delta\mu$  for  which  the
conduction band  is always inside  the populated region.   In the
rest  of our  numerical examples  we set  $U =  0$ and  $t  = 1$.
Analogously, in the right panel of Fig.~\ref{j-t} we show that for fixed
$t$ the current grows  linearly with $\Delta\mu$ and saturates at
$\Delta\mu >  4t$.

\begin{figure}
 \includegraphics[width=7cm]{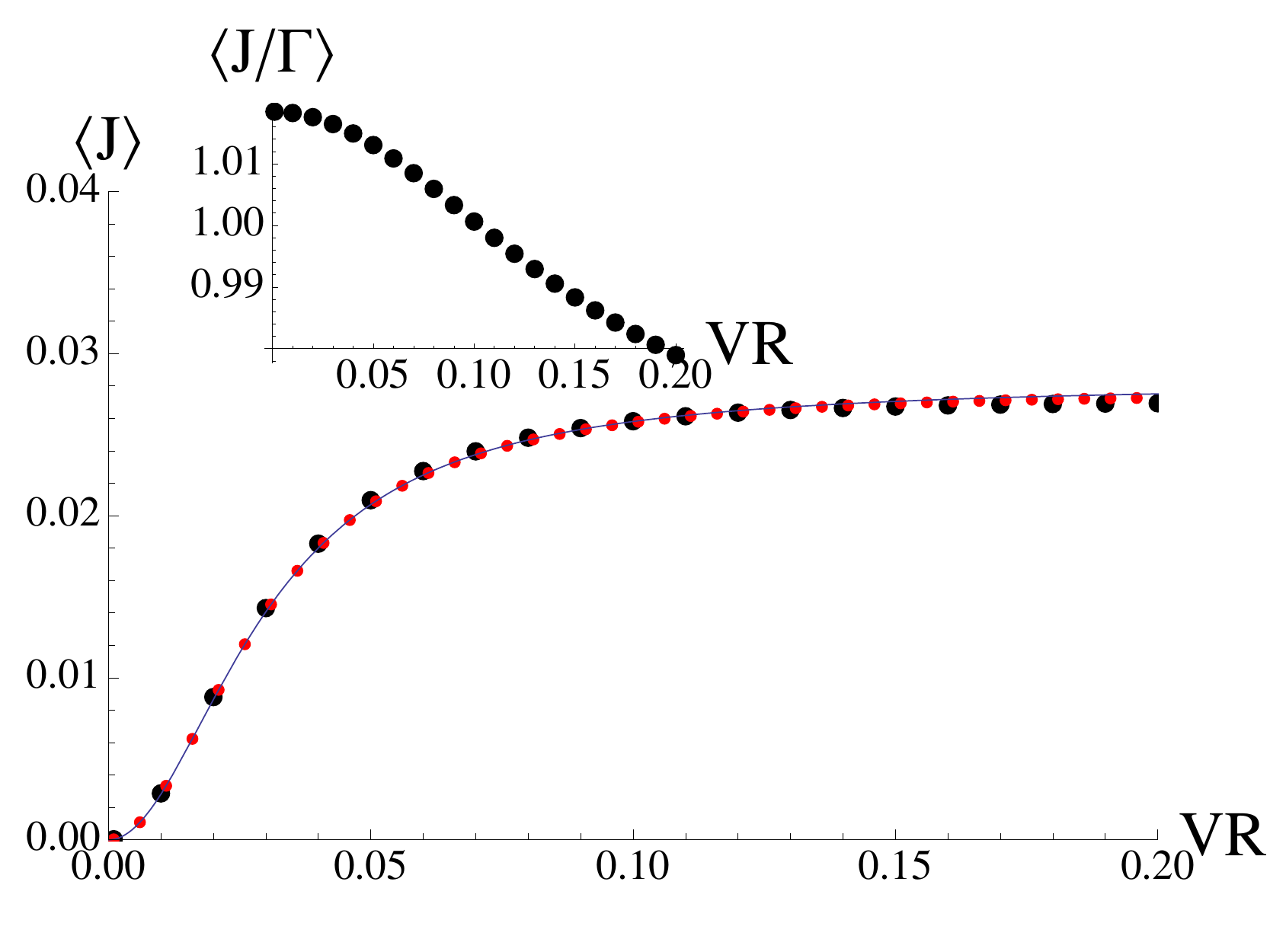}
\caption{Black points are obtained by our model with $\gamma = 0.1, \,\beta_L = \beta_R=5,\, \mu_R = -4,\, \mu_L =4;
N= 10,\,U = 0,\, t = 1,\, v_L = 0.03,\,K=200$ and $E_{\rm min}=-20,\,E_{\rm max}=20$. The red dots are obtained by a numerical computation of the Landauer formula with $T=0$, and the continuous line is $\ave{J}=\Gamma=2\pi\frac{v_R^2v_L^2}{\theta_0(v_L^2+v_R^2)}$. The inset is $\ave{J}/\Gamma$. }
\label{j-vr}
\end{figure}

What is  perhaps more interesting  is the scaling of  the current
with $\Gamma=\Gamma_L\Gamma_R/(\Gamma_L+\Gamma_R)$.  In Fig.~\ref{j-vr}  we
plot the current  $\ave{J}$ as a function of the  coupling to the right
lead,  showing that  the main  trend of  the current  is $\langle
J\rangle  =  \Gamma$.   In  the  inset we  show  that  there  are
deviations to this law.

We have analyzed this behavior using the Landauer formula, that allows deeper analytical exploration. Since temperature is very low we take it exactly zero in both reservoirs, thus the Landauer formula is 
$\ave{J}=\Gamma_L\Gamma_R/2\pi \int_{\mu_L}^{\mu_R} {\rm tr}[G^+ (\omega) G^-(\omega)].$
Changing the representation of the Green matrix from position basis to the energy basis (of the isolated chain) one can see that it is possible to approximate (isolated resonances approximation) the integrand as a
superposition of $N$ Lorentzians each located at $2t\cos[ j\pi/(N + 1)]$ and with a width $(\Gamma_L +\Gamma_R) \sin^2 [j \pi/(N + 1)]/(N+ 1), \, j=1,\ldots N$ along the energy axis. If $t<\mu_L$, the integral can be extended from $-\infty$ to $\infty$ because the Green function decays exponentially outside the energy band of the chain, thus the result $\ave{J}=\Gamma$ is obtained. 

\begin{figure}
 \includegraphics[width=7cm]{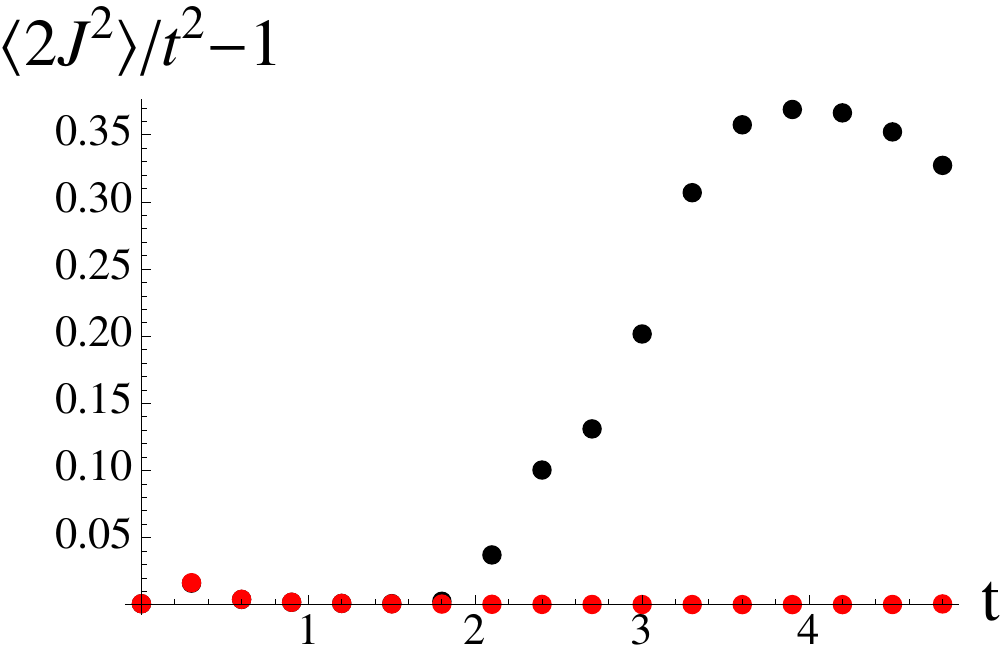}
 \includegraphics[width=7cm]{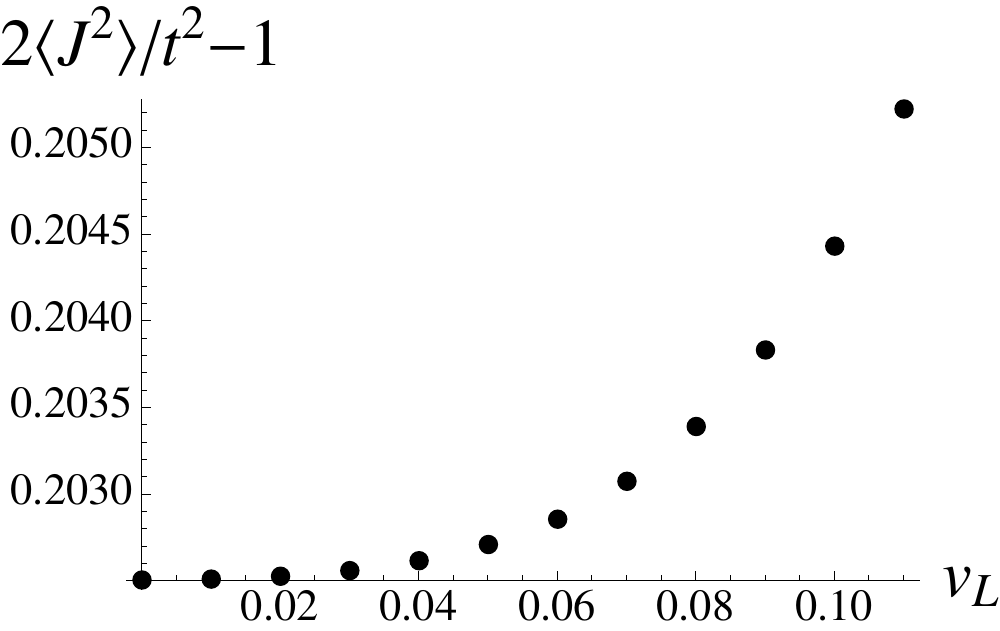}

\caption{Panel a: $2\ave{J^2}/t^2-1$ as function of $t$ for $\mu_R = -4,\, \mu_L =4$ (black dots) and $\mu_R = -10,\, \mu_L =10$ (red dots) as in Fig. \ref{j-t}a.  Panel b: $2\ave{J^2}/t^2-1$ for $t=3$ as a function of $v_L=v_R$. In both figures the other parameters are: $\gamma = 0.1, \,\beta_L = \beta_R=5,
N= 10,\,U = 0,\,K=200$ and $E_{\rm min}=-20,\,E_{\rm max}=20$ with $v_L = 0.03$ in panel a and $t=3,\mu_R = -4,\mu_L =4$ in panel b. 
}
\label{fluct}
\end{figure}

Now we can also explore fluctuations and compute $\ave{J^2}$. In Fig.\ref{fluct}a, we computed for the same parameters than in Fig.\ref{j-t}a the quantity 
$2\ave{J^2}/t^2-1$. We see that as soon as the current decreases because some modes of the chain goes out of the populated energy region, the fluctuation increases. As a function of other parameters like $v_\alpha$ or $N$ (data not shown), the average $\ave{J^2}$ does not have important changes (see Fig.\ref{fluct}b).
\section{conclusions}
\label{conclu}

We showed that in the wide-band limit, the numerical results found in our model indeed correspond to what is expected on the basis of the Landauer formula, a formula which is usually interpreted as if the reservoirs were always in an equilibrium (gran canonical) distribution, not perturbed by the presence of the system. The Landauer formula emphasizes the role of the Fermi distributions of the reservoirs and provides an accurate description of the current if the assumption of reflectionless contacts is justified. In this respect, 
a very interesting relation was found in~\cite{Shigeru} and proved in~\cite{mariborg} between the current and the occupation of the meso-reservoir 
$\ave{J} =\sum_{k=1}^K2\gamma_k[\langle n^L_k\rangle-F_L(\varepsilon_k)]$. This is an exact relation that in the appropriate limit should converge to the Landauer formula. Note that it implies that the occupation difference with respect to the Fermi distribution is 
${\mathcal O}(\Gamma/\theta_0\gamma)$. 
 It is a very interesting relation because it links the current, which is the fingerprint of the non equilibrium state, to the difference in distribution to the equilibrium case. Something similar has been found in classical systems where the fractal nature of the non equilibrium state is determined by the current~\cite{gaspard}. 
Moreover in ~\cite{Shigeru} we analyzed how Onsager reciprocity relation is broken in the system and found that $|L_{up}/L_{pu}-1|$ grows with $\gamma$ implying that despite the  almost $\gamma$ independent value of the current, the dissipative mechanisms in the super-reservoir play an important role. 
A deeper study of these effects, that are beyond the Landauer picture, can be studied in the context of the model presented here and deserve further investigation.

\acknowledgments

The authors gratefully acknowledge the hospitality of NORDITA
where part of this work has been done during their stay, within
the framework of the Nonequilibrium Statistical Mechanics
program, and support from ESF.
FB thanks Fondecyt project 1110144 and SA  thanks Fondecyt project 3120254.


\begin{thebibliography}{99}
\bibitem{needref} W Aschbacher, V Ja\v{k}si\'c, Y Pautrat, and C Pillet, J. Math. Phys. {\bf 48}, 032101 (2007). 
\bibitem{Shigeru} S. Ajisaka, F. Barra, C.Mejia-Monasterio and T. Prosen arXiv:1204.1321
\bibitem{mariborg} S. Ajisaka, F. Barra, C.Mejia-Monasterio and T. Prosen, arXiv:1205.1167.
\bibitem{Rue00} D. Ruelle J. Stat. Phys. {\bf 98}, 57 (2000); Comm. Math. Phys. {\bf 224}, 3 (2001).
\bibitem{tasaki} S. Ajisaka, S. Tasaki and F. Barra, Bussei-Kenkyu {\bf 97} 483 (2011/2012); arXiv:1110.6433
\bibitem{Wichterich07} H. Wichterich, M. Henrich, H-P. Breuer, J. Gemmer, and M. Michel, Phys. Rev. E {\bf 76}, 31115 (2007).
\bibitem{prosen08} T. Prosen, N. J. Phys. {\bf 10}, 3026 (2008).
\bibitem{Kosov11} A. A. Dzhioev and D. S. Kosov, J. Chem. Phys.  044121 {\bf 134} (2011).
\bibitem{Prosen10} T. Prosen J. Stat. Mech. P077020 (2010).
\bibitem{book} S. Datta, {\em Electronic Transport in Mesoscopic Systems} (Cambridge University Press, Cambridge, 1995).
\bibitem{silly} M. Zilly, {\em Electronic conduction in linear quantum systems: Coherent transport and the effects of decoherence}  PhD. dissertation Universit\"{a}t Duisburg-Essen (http://duepublico.uni-duisburg-essen.de/servlets/DerivateServlet/Derivate-24279/dissertation\_zilly.pdf).
\bibitem{felipe} F. Barra and P. Gaspard, J. Phys. A Math and Gen. {\bf 32} 3357 (1999). 
\bibitem{gaspard}P. Gaspard, {\em Chaos, Scattering and Statistical Mechanics} (Cambridge University Press, Cambridge, 1998).
\end{thebibliography}
\end{document}